\documentstyle[aps,prl]{revtex}
\input{epsf}
\begin{document}
\twocolumn[\hsize\textwidth\columnwidth\hsize\csname@twocolumnfalse\endcsname
\bibliographystyle{unsrt}

\title{Anomalous relaxation and  self-organization in non-equilibrium
processes}

\author{Ibrahim Fatkullin $^1$,  Konstantin Kladko $^2$, Igor Mitkov $^3$ and 
A.R. Bishop $^4$}

\address{$^1$Department of Mathematical Sciences, Rensselaer Polytechnic 
Institute,
Troy, NY 12180 \\
$^2$Department of Physics, Stanford University,
Stanford, CA 94305
\\
$^3$Department of Physics and CIRCS, Northeastern University,
Boston, MA 02115 \\
$^4$Theoretical Division and Center for Nonlinear Studies,
Los Alamos National Laboratory, Los Alamos, NM 87545 
}
\date{\today}
\maketitle

\begin{abstract}
We study thermal relaxation in ordered arrays of coupled nonlinear 
elements with external driving. We find, that our model exhibits
dynamic self-organization
manifested in a universal stretched-exponential
form of relaxation. We identify two types of self-organization,
cooperative and anti-cooperative, which lead 
to fast and slow relaxation, respectively. We give a qualitative
explanation for the behavior of the stretched exponent in 
different parameter ranges. We emphasize that this is
a system exhibiting stretched-exponential relaxation without
explicit disorder or frustration.
\end{abstract}

%
% \pacs{PACS: ???????????}

\narrowtext
\vskip1pc]

The world around us is full of non-equilibrium and non-stationary
processes, many of which are robust and easy to measure. However,
there is no
a priori reason to believe, that these non-equilibrium processes
should fulfill universal laws, in a way equilibrium systems fulfill
laws of thermodynamics and statistical mechanics. It is, therefore,
important to recognize examples of complex non-equilibrium processes
organizing themselves in a simple and universal way, a phenomenon
termed {\it ``self-organization''} or {\it ``emergent behavior''} 
\cite{laughlin1,laughlin2}.
Kolmogorov turbulence~\cite{kolmogorov} is an example of such
non-equilibrium process: Liquid, forced at macroscopic length scales, 
produces a flow of energy from large to small length scales, and this
flow organizes itself into a stationary universal distribution. 
Turbulence is a steady, though non-equilibrium,
process; the flow of energy is constant and the resulting
self-organized probability distribution does not change in time. 
It is important to establish, whether boundaries of self-organization
can be expanded to include non-equilibrium and non-stationary processes.
In this paper we show that {\it dynamic self-organization} can indeed
be found in relaxational dynamics
of extended systems and is manifested in stretched-exponential 
dependence on time of physical quantities.
   
Stretched-exponential
(\mbox{$\propto \exp \left[ -(t/t_{rel})^\alpha \right]$}) 
relaxation laws have been observed in a large variety of physical and
biological processes, such as recombination of carriers in semiconductors
and polymers~\cite{semiconductors,polymers}, protein
relaxation~\cite{myoglobinfolding} and folding~\cite{downhill,frauenfelder2}, 
ligand binding to myoglobin~\cite{myoglobin,frauenfelder}, relaxation  
in magnetic clusters~\cite{kent}, superconducting
vortices~\cite{superconductors} and charge density waves~\cite{cdw1,cdw},
dynamics of alloys~\cite{alloys} and glasses~\cite{glasses}. 
In the case of glasses, the stretching exponent $\alpha$
defines a glass transition temperature, $T_g$, {\it i.e.} \mbox{$\alpha=1$}
for \mbox{$T>T_g$}, and  \mbox{$\alpha<1$} for \mbox{$T<T_g$}.
The non-exponential dynamics of glasses has long been related
to the high degree of disorder, which leads to the existence
of a large number of metastable states~\cite{anderson}.

Observations of non-exponential
behavior in simpler systems, such as magnetic clusters~\cite{kent}  
and proteins~\cite{myoglobinfolding,downhill}, suggests that a high degree
of disorder is {\it not} a necessary requirement for
a system to display anomalous relaxation. Especially notable are recent
observations of folding dynamics in proteins, yeast
phosphoglycerate kinase (PGK) and ubiquitin mutant~\cite{downhill}.
These proteins
fold according to a ``downhill folding'' scenario~\cite{eaton},
meaning that the folding path between the unfolded and folded
states is free of deep metastable minima, so the process of folding
is ``downhill'' relaxation along this path. Nevertheless, the number
of folded proteins, as a
function of time, displays stretched-exponential behavior over a
large time interval~\cite{downhill}.

In this Letter we present a simple minimal model which is
translationally invariant and {\it without} disorder or explicit
frustration~\cite{bishop}, which displays perfect stretched-exponential
relaxation over wide time intervals.
The stretched exponent $\alpha$ changes continuously from slow relaxation
\mbox{$\alpha<1$} to fast relaxation \mbox{$1 \le \alpha \le 2$}, as  
parameters of the system vary.
We identify {\it dynamic self-organization}
as the origin of the stretched-exponential relaxation, and show, that
slow and fast relaxation
are caused by {\it anti-cooperative} and {\it cooperative} behavior, 
respectively. We provide a theoretical explanation of the discrete and
continuous limiting cases. We also present a qualitative theory, which
accounts for behavior of the stretched exponent in
the intermediate range of parameters.

Real life systems [4-9] are much more complex than our
simple model. However, we conjecture that at least one underlying
reason for stretched-exponential relaxation is universal. Namely,
{\it self-organization}, with fast and slow relaxation corresponding
to cooperative and anti-cooperative behavior. We suggest
that proteins belong to the anti-cooperative universality class. Our
argument is, that initially the protein is loose   
and deforms easily. When folding into a particular local pattern occurs,
this part becomes stiff, making folding of neighboring 
parts more difficult, and therefore behaving in an anti-cooperative way. 
Similar anti-cooperative behavior occurs in our model system for
negative values of the coupling, as we discuss below.

Let us now specify the model. We consider a chain of nonlinear bistable
elements. Each element is described by the order parameter $u_n$. 
The local energy $E(u)$ has two minima, one of which is a 
metastable state (local minimum of energy), the other one is the absolutely
stable state (absolute minimum of energy). We assume overdamped dynamics of
$u_n$ at the presence of delta-correlated thermal noise. 
Equations of motion of the system are:

\begin{equation}\label{main}
\frac{\partial u_n}{\partial t}=\beta (u_{n+1} + u_{n-1} -2 u_n) 
+F(u_n)+f_n^{\,\rm{stc}}.
\end{equation}

Here $\beta$ represents linear coupling of neighboring sites,   
$F(u) = -dE/du = -u(u-u_0)(u-1)$ is a forcing term, corresponding
to  the bistable polynomial potential
$E(u) = (1/4) u^4- (1/3)(u_0+1) u^3 + (1/2) u_0 u^2 $.
The potential has two minima,
$u=0$ and $u=1$, separated by a barrier, $u=u_0$. At $u_0=1/2$ 
the minima have equal energy. For $u_0 < 1/2$ the minimum  $u=1$ becomes 
absolutely
stable and $u=0$ metastable, with  energy difference  given by
$\Delta E = 1/12 -1/6 u_0$. The stochastic term
$f^{{\,\rm stc}}$ is given by a delta-correlated Langevin force
\begin{equation}
\label{deltacorrelated}
        <f^{{\,\rm stc}}_n(t_1)f^{{\,\rm stc}}_m(t_2)>=
        T\:\delta_{mn}\delta(t_1-t_2),
\end{equation}
where $T$ defines the temperature in our system~\cite{vankampen}. 
Eq.~(\ref{main}) can be viewed 
as discrete one-dimensional Ginzburg-Landau equation with noise.

\begin{figure}[h]
\hspace{-1cm}
\rightline{ \epsfxsize = 8.5cm \epsffile{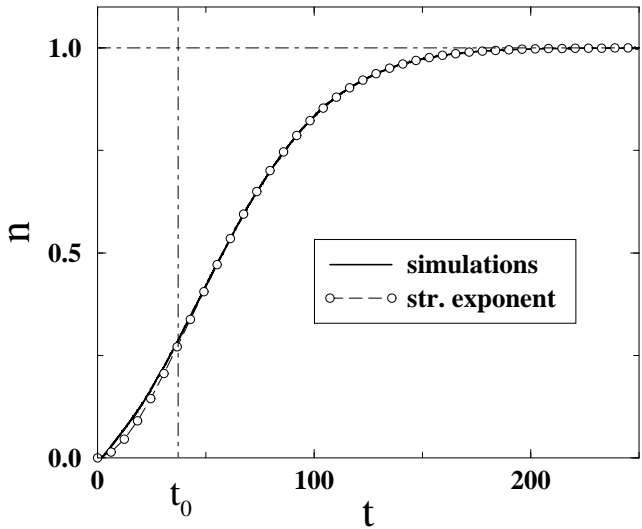}}
\caption{
Plot of a typical function $n(t)$. Solid line represents the
result of numerical simulations. Dashed line is a stretched-exponential
fit. Parameters are $T=0.01$, $u_0= 0.35$, $\beta=0.04$.
\label{fig1}}
\end{figure}

Let us first review equilibrium properties of our system. At $T=0$
the system is in the absolutely stable phase. When temperature
is introduced, some (small) number of particles overcome the barrier
because of equilibrium thermal fluctuations. The relative number of
particles in the metastable phase is determined by Boltzmann statistics,
and is exponentially small for temperatures
much less than the energy difference between potential wells. 

Our goal is to study non-equilibrium and non-stationary process of
relaxation from the metastable phase to the absolutely stable phase.   
We assumes that all sites are initially in the metastable phase, and
then introduces  the  thermal noise. Due to fluctuations, particles
start to overcome the barrier, and the number of particles in the absolutely 
stable phase increases. We monitor relaxation by introducing the function
$n(t)= N(t)/N_0(T)$, which describes the ratio between the
concentration of particles in the absolutely stable phase at time $t$
and their equilibrium number $N_0(T)$. In the thermodynamic limit
of infinite chain length, $n(t)$ is a well defined smooth function, which
satisfies the conditions $n(0)=0$ and $n(\infty)=1$.   

We study  relaxation numerically by integrating the Langevin equations
(\ref{main}) defined on long chain segments ($N=200$), and averaging
the resulting functions $n(t)$ over many noise realizations. To improve
numerical convergence, we use an implicit integration scheme. Details of
the numerical method will be reported elsewhere~\cite{longpaper}.  

\begin{figure}[h]
\hspace{-1cm}
\rightline{ \epsfxsize = 8.5cm \epsffile{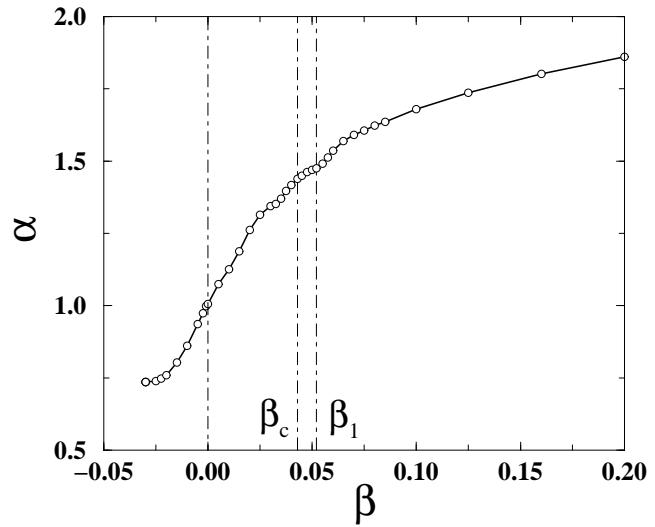}}
\caption{
Stretched exponent $\alpha$ as a function of inter-site coupling $\beta$,
for $u_0$ and $T$ the same as in Fig.~\protect\ref{fig1}. Propagation
failure bifurcation and one-site
nucleus bifurcation are marked as $\beta_c$ and $\beta_1$, respectively.
\label{fig2}}
\end{figure}

\begin{figure}[h]
\hspace{-1cm}
\rightline{ \epsfxsize = 8.5cm \epsffile{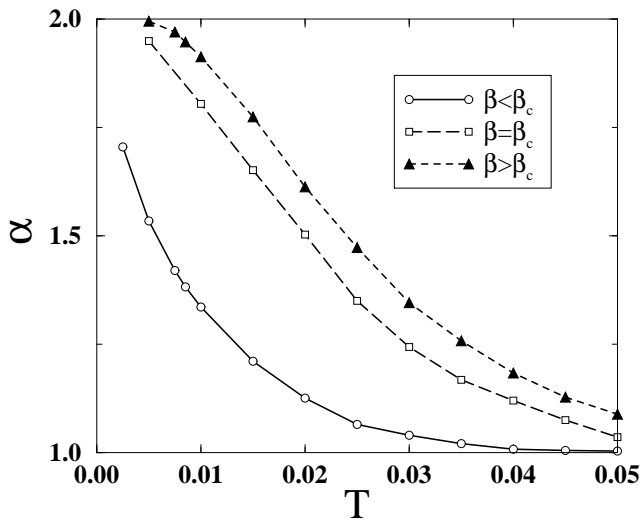}}
\caption{
Stretched exponent $\alpha$ as a function of temperature  $T$,
for three different values of $\beta$, above, below, and at the 
propagation failure bifurcation point. Parameters are $u_0=0.35$,
$\beta=0.03;0.043;0.06$. Other parameters are the same as in
Fig.~\protect\ref{fig1}.
\label{fig3}}
\end{figure}

A typical function $n(t)$ is given in Fig.~\ref{fig1}. 
Important global features of the relaxation,
valid in all ranges of parameters, are:

i) After a short transient time $t_0$ (Fig.~\ref{fig1}), the system
self-organizes and $n(t)$ starts to obey the
stretched-exponential form $n(t) = 1 - \exp[-(t/t_{rel})^{\alpha}]$. 
Note, that there are only two parameters in our fit, which are the
stretched exponent $\alpha$ and the relaxation time $t_{rel}$.
The precision of the stretched-exponential fit is extremely high,
and is about $0.1\%$ for $t>t_0$;

ii) The stretched exponent $\alpha$ changes continuously,
as temperature $T$ and coupling $\beta$ vary, see
(Figs.~\ref{fig2} and \ref{fig3}).
The observed relaxation is fast ($\alpha>1$) for positive values of
$\beta$ and slow ($\alpha<1$) for negative values of $\beta$.
As temperature $T$ is increased, the stretched exponent $\alpha$ 
approaches the Arrhenius law ($\alpha=1$), corresponding to
the intrinsic frustration being overcome thermally;
    
iii) The stretched exponent $\alpha$ approaches the value $\alpha=2$
in the continuous limit of large $\beta$, representing a binary
relaxation channel.

The dependence of $\alpha$ on $\beta$ is given in Fig.~\ref{fig2}.
At zero $\beta$ the system represents a set of uncoupled nonlinear
sites. Each site is described by a one-dimensional Fokker-Planck
equation. In this case relaxation is known to be described 
by an exponential (Arrhenius) law, with  decay time determined by
the lowest excited state of the Fokker-Planck operator~\cite{vankampen}. 
Therefore,
for $\beta=0$ one has $\alpha=1$.

Let us now explain, why introducing positive/negative $\beta$ corresponds
to  cooperative/anti-cooperative behavior, respectively.
If $\beta$ is positive, one may think of the inter-site coupling $\beta$
as an elastic chain connecting neighboring sites $n$ and $n+1$.  
If a particle overcomes the barrier, it attracts its neighbors and makes
it, therefore, easier for them to jump over the barrier. Contrary
to this, negative coupling corresponds to a repulsive force, which  
makes relaxation of nearest neighbors more difficult. (Indeed in
the limit of large negative $\beta$, our model produces a strong
staggered dimerization of the lattice). We conclude from
Fig.~\ref{fig2} that the cooperative behavior leads to fast relaxation,
$\alpha>1$, and anti-cooperative behavior leads to slow relaxation,
$\alpha<1$.

%We conjecture, that this correspondence of 
%cooperative/anti-cooperative
%behavior to fast/slow relaxation is valid for a large class of systems, and,
%in particular, for proteins.

\begin{figure}[h]
\hspace{-1cm}
\rightline{ \epsfxsize = 8.5cm \epsffile{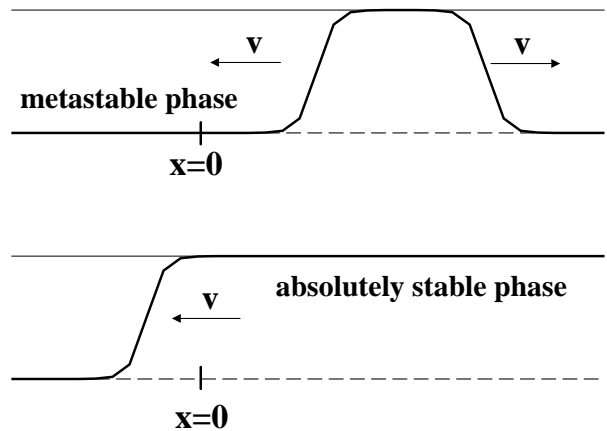}}
\caption{
New phase nucleation through birth and propagation of kink-antikink pairs
(schematic).}
\label{fig4}
\end{figure}

We will now discuss the behavior of the stretched exponent $\alpha$ 
at large positive $\beta$. In this limit the system becomes continuous and
is described by the continuous Ginzburg-Landau equation.
\begin{equation}\label{continuous}
\frac{\partial u}{\partial t}=\beta \frac{\partial^2 u}{\partial x^2}{}  
+F(u)+f(x,t)^{\,\rm{stc}}.
\end{equation}
In the absence of thermal noise, topological excitations of this 
equation are fronts (kinks), which separate the absolutely stable and 
metastable phases. Due to the energy difference between the phases,
the fronts propagate at finite velocity $v$, increasing the size of
the absolutely stable phase.
When temperature is introduced, local fluctuations of the order parameter
give birth to kink-antikink pairs (Fig.~\ref{fig4}). These pairs 
counter-propagate, replacing the metastable phase by the absolutely stable 
one. Let us now estimate the probability for the order parameter
$u(0,t)$ at $x=0$ to stay in the metastable phase after time $t$. This
probability is approximately equal to the probability $P(t,l)$, that
no kink-antikink pair will be created by fluctuations during time $t$
at a distance $l \lesssim vt$ from the origin.
If such a pair is created, then the newly-born kink has enough
time to reach $x=0$, before the time interval $t$ elapses, and annihilate
the metastable phase; see Fig.~\ref{fig4}. Since fluctuations are local,
one can estimate $P(l,t)$ as $\exp[-lt/\eta]$, where $\eta$ is a constant.
Therefore, the probability to stay in the metastable phase after time $t$
is approximately $\exp[-vt^2/\eta]$. Since the number of particles in
the metastable phase after time $t$ is proportional to this probability,
we conclude that in the continuous limit the stretched exponent
$\alpha=2$. Our numerical data (see Fig.~\ref{fig2}) are in a good
agreement with this prediction.

The fact, that the stretched exponent $\alpha=2$ in the continuous case,
is striking and has important consequences. Namely, {\it space
discreteness is necessary to observe continuous dependence of 
$\alpha$ on parameters}. We can rephrase this statement as follows. 
It is known, that critical properties of equilibrium systems are
described by continuous models, and these models can be organized
into universality classes, such that each universality class has 
a set of universal critical exponents~\cite{wilson,zinnjustin}. We 
conjecture, that effective models for non-equilibrium, non-stationary
self-organization, in particular relaxation, must include discreteness,
in order to incorporate the continuous dependence of stretched exponents
on parameters.

Let us now consider the nontrivial dependence of
the stretched exponent $\alpha$ on the coupling $\beta$ and
temperature $T$, for the intermediate values of $\beta$, given in
Figs.~\ref{fig2} and \ref{fig3} (a more detailed discussion of
these effects will be given in~\cite{longpaper}).
In the absence of noise, the topological excitations of 
system~(\ref{main}) undergo a sequence of bifurcations,  
as described in~\cite{pearson,igor,kladko}. The most 
important of these bifurcations is the propagation failure bifurcation,
at which fronts cease to propagate,
being pinned by the lattice. We denote the corresponding value of
the coupling $\beta$ as $\beta_c$. An analytical expression for
$\beta_c$ was derived by us in~\cite{kladko}.
The second important bifurcation point is $\beta_1$. At this
point, a one-site nucleus of the globally stable phase, representing
a bound state of a kink and antikink pair, becomes unstable, and
''bursts'' into counter-propagating kink and antikink. In~\cite{igor}
we have shown, that there exists an infinite number of bifurcation
points $\beta_n$, such that $\beta_c \le ... \le
\beta_n \le \beta_{n-1}  ... \le \beta_1$, corresponding to the
instabilities of a one-site, two-site, three-site, etc. nuclei.  
In Fig~\ref{fig2} we see the corresponding change of shape in
the coupling-dependence of the stretched exponent $\alpha$,
in the interval $\beta_c < \beta < \beta_1\,$. Accordingly,
Fig.~\ref{fig3} demonstrates a transition in the
temperature-dependence of $\alpha$, with $\beta$ crossing $\beta_c$.
We conclude from Figs.~\ref{fig2} and \ref{fig3}, that this region
of bifurcations is manifested as a characteristic feature of the
non-equilibrium relaxation of the system.

In conclusion, we have introduced a model of dynamic self-organization,
and conjectured that it should be applicable to stretched-exponential
relaxation in many biological and physical systems. We studied properties
of self-organization in a simple model, and showed that fast/slow
relaxation is related to cooperative/anti-cooperative behavior. 
We showed, that neither explicit disorder nor frustration are necessary
requirements for anomalous relaxation laws. In view of this, an
experimental evidence of anomalous relaxation in a physical system
should not be considered as automatic proof of disorder or
frustration in the system.
Appropriate nonlinearity (arising, for example, from coupled
degrees-of-freedom with feedback) can alone produce both the
ingredients of local structures (''intrinsic disorder'') and
competition (of length-scales), leading to a distribution of
metastable states and anomalous dynamics~\cite{bishop}.
The detailed  analytical description  
of dynamic self-organization is a promising direction for
future study. This will include extensions to higher
dimensions as well as specific physical models.
We also note, that stretched exponents were also observed in 
{\it equilibrium} correlation functions~\cite{flach}, and it is 
important to understand, whether they can be related to their   
non-equilibrium counterparts.

We thank R. B. Laughlin for fruitful discussions and for sharing
his views on self-organization in non-equilibrium processes.
We are also grateful to J. Pearson, for his valuable comments.
This work was supported in part (K.K.) by the Otto Hahn Fellowship of 
the Max Planck Society, Germany. Work at Los Alamos was supported by
US DOE under contract No. W-7405-ENG-36.

\bibliography{selforg}

\end{document}